# Physics-based model to predict the acoustic detection distance of terrestrial autonomous recording units over the diel cycle and across seasons: insights from an Alpine and a Neotropical forest

Sylvain Haupert [a,*], Frédéric Sèbe [b], Jérôme Sueur [a]

[a] Muséum national d'Histoire naturelle, CNRS UMR 7205, ISYEB, Sorbonne Université, Paris, France
[b] ENES Bioacoustics Research Laboratory, University of Saint-Etienne, CRNL, CNRS UMR 5292, Inserm UMR_S 1028, Saint-Etienne, France

[*] Corresponding author: Institut de Systématique, Évolution, Biodiversité (ISYEB), Muséum national d'Histoire naturelle, CNRS, Sorbonne Université, EPHE, 57 rue Cuvier, 75005 Paris, France, +33 608101192, sylvain.haupert@mnhn.fr

**Abstract**

1. Passive acoustic monitoring of biodiversity is growing fast, as it offers an alternative to traditional aural point count surveys, with the possibility to deploy long-term acoustic surveys in large and complex natural environments. However, there is still a clear need to evaluate how the frequency- and distance-dependent attenuation of sound as well as the ambient sound level impact the acoustic detection distance of the soniferous species in natural environments over the diel cycles and across seasons. This is of great importance to avoid pseudoreplication and to provide relevant biodiversity indicators, including species richness, species abundance and species density.

2. To address the issue of detection distance, we tested a field-based protocol in a Neotropical rainforest (French Guiana, France) and in an Alpine coniferous forest (Jura, France). This standardized and repeatable method consists in a recording session of the ambient sound directly followed by an experiment using a calibrated white noise sound broadcast at different positions along a 100 m linear transect. We then used acoustic laws to reveal the basic physics behind sound propagation attenuation.

3. We demonstrate that habitat attenuation in two different kinds of forests can be modelled by an exponential decay law with a linear dependence on frequency and distance. We also report that habitat attenuation, as first approximation, can be summarized by a single value, the coefficient of attenuation of the habitat.

4. Finally, we show that the detection distance can be predicted knowing the contribution of each attenuation factor, the coefficient of attenuation of the habitat, the ambient sound pressure level and the amplitude and frequency bandwidth characteristics of the transmitted sound. We show that the detection distance mostly depends on the ambient sound and may vary by a factor of up to 5 over the diel cycle and across seasons. These results reinforce the need to take into account the variation of the detection distance when performing passive acoustic surveys and producing reliable biodiversity indicators.

1. **Introduction**

Passive acoustics monitoring is becoming an attractive sampling tool in ecology, in particular for biodiversity monitoring, as it provides complementary data to traditional aural point count surveys (Gibb et al., 2019) and camera trap surveys (Buxton et al., 2018). Remote acoustic sensing devices, also called ***autonomous recording units*** (ARUs), offer the possibility to record soundscapes in terrestrial (Sugai et al., 2019), marine (Sousa-Lima et al., 2013) and freshwater environments (Desjonquères et al., 2020), for long periods (Folliot et al., 2022; Grinfeder et al., 2022). ARUs increase the sampling effort and at a frequency range that can exceed the human ear range, e.g. infrasound (Fregosi et al., 2020) or ultrasound (Newson et al., 2017), with relatively low-cost devices (Hill et al., 2019). In addition, the soundscapes recorded by ARUs are analyzed *a posteriori*, offering multiple types of repeatable analyses based on human experts (Yip & Bayne, 2017) and/or algorithms (Ulloa et al., 2018).

Recent advances in the automatic recognition of species vocalizations (Kahl et al., 2021) open the possibility of detecting the presence of target species in soundscapes and thereby attaining important information for biodiversity assessment and conservation (Priyadarshani et al., 2018). To deliver reliable species richness or abundance data at specific sites and to assign a specific number of observations per unit area, it is often necessary to estimate the detection distance of an ARU, although this is rarely done in practice (Pérez-Granados & Traba, 2021), probably because it requires the realization of a difficult experiment without any standardized guidelines (Sugai et al., 2020). Furthermore, knowing the detection distance is a prerequisite for defining the number and position of ARUs that need to be installed in the field to properly cover a study area (Pérez-Granados et al., 2018). Standardization in the evaluation of the detection range is also required to provide correction factors between different brands of ARUs (Yip et al., 2017) as well as between ARUs and human observers (Castro et al., 2019; Darras et al., 2018, 2019; Van Wilgenburg et al., 2017; Yip et al., 2017). Therefore, determining the detection distance, also referred to as the sampling area, effective detection radius or detection range, appears to be key to obtaining reliable biodiversity estimations (Shonfield & Bayne, 2017).

Most acoustic surveys estimate the detection distance by determining the threshold distance from which the sound produced by a species of interest is no longer recognizable (Yip et al., 2017) or is below the ambient sound pressure level (Darras et al., 2016). This estimation is generally achieved by playing back focal species-specific song/call or tone bursts at different distances from the recorder (Shaw et al., 2021). Statistical models such as the well accepted effective detection radius (EDR) (Sólymos et al., 2013) or the recent distance truncation method (Hedley et al., 2021) are then used to determine when a sound is no longer recognizable by an algorithm or a human expert. Such models allow multiple factors to be tested, such as the importance of ARU specifications (e.g. sensitivity, frequency response), sound source properties (e.g. sound levels, pure tones, species vocalizations, speaker brand, height or orientation), habitat characteristics (e.g. deciduous, coniferous, open or closed habitat, ambient sound) and environmental conditions (e.g. relative humidity, temperature, wind speed) (Darras et al., 2016, 2018; MacLaren et al., 2018; Pérez Granados et al., 2019; Priyadarshani et al., 2018; Shaw et al., 2021; Thomas et al., 2020; Yip et al., 2017; Yip & Bayne, 2017). Despite these models being widely used, they suffer from some drawbacks. These models are mostly species-specific with little consideration for the frequency characteristics of the animal vocalizations, precluding any generalizations to other species with different frequency characteristics. In addition, these models do not consider the underlying physics of sound propagation – such as the relative role of sound atmospheric absorption and sound scattering due to vegetation – which is required to simulate realistic animal vocalization attenuation after propagation. Finally, these models do not directly assess signal masking by the ambient sound as well as its variation over the diel cycle and across seasons which might be an important driver of the detection range.

In this article, we propose to address some of the aforestated limitations by introducing a standardized, repeatable and generalized method that is able to predict the ***attenuation*** of sound depending on the physical laws that govern the sound propagation in natural environments. This new approach, which is complementary to the aforementioned statistical models, is based on the popular ***excess attenuation*** framework of outdoor sound propagation, which estimates the surplus of attenuation that is not explained by the sum of the well known ***geometric*** and ***atmospheric attenuations*** (Aylor, 1972; Price et al., 1988). After carefully removing the ***ambient sound*** contribution from the audio recordings, we were able to estimate the ***frequency***- and ***distance***-dependent coefficient of attenuation. This is a key parameter to simulate sound attenuation in two closed but contrasting habitats, a Neotropical rainforest (French Guiana, France) and an Alpine coniferous forest (Jura, France). We determined the ***detection distance*** of an ARU corresponding to the ***distance*** at which the ***sound pressure level*** of a transmitted signal goes below the ambient sound. Finally, we modelled the variation in the detection distance of the ARU over the diel cycle and across seasons in both habitats in order to analyze its temporal fluctuations and its impact on long-term monitoring. Explanations of technical terms related to physical acoustics used in this paper are given in the supplementary glossary and their first relevant use in the text is indicated in bold italics.

## 2. Materials and Methods

### a. Study sites

We performed propagation experiments in two forest sites. The first site was a pristine Neotropical lowland rainforest (French Guiana, France), near the Saut Pararé rapids of the Arataye River (4°2'N; 52°40'W, CNRS Nouragues Research Station). The forest is characterized by a fairly open understorey and a dense canopy between 40 m and 45 m (see Table A-1 of Appendix A). The ground is covered with thick litter. The habitat structure does not undergo cyclic nor important changes over the year. The experiment was conducted on 18 February 2019, at the end of the dry season. The temperature and the relative humidity during the propagation test were 27.9 °C and 87 %, respectively. The atmospheric pressure given by the nearest Météo France station at Cayenne-Matoury (97 km away) was 101,340 Pa.

The second site was an Alpine coniferous forest located in the Massif du Risoux (46°32'N; 6°06'E), within the Parc Naturel Régional du Haut-Jura (Jura, France). The site is mainly covered by pure spruce (*Picea abies*) forest (see Table A-2 of Appendix A). The limestone soil is composed of a thin litter, covered with blueberries (*Vaccinium myrtillus*) and presenting a superficial groove carved out by water in limestone soil called lapiaz. The habitat structure does undergo small changes over the year, with snow cover during the winter and slight changes of the foliage density due to the presence of some deciduous trees and shrubs. The experiment took place on 11 July 2019, with an average temperature of 17 °C and a relative humidity of 67 %. The atmospheric pressure given by the nearest Météo France station at Dijon-Longvic (114 km away) was 87,999 Pa.

Finally, we used one-year audio data sets collected by the same ARU (Song Meter SM4, Wildlife Acoustics, Maynard, MA, USA) with the same audio configuration at each site. These data sets consisted in a collection of 1-minute audio recordings every 15 minutes (1 minute on, 14 minutes off) all day long for a total of 584 hours (96 minutes * 365 days). These data sets were used to estimate the variation in the detection distance during the night and day cycle and across seasons. In both cases, the relative humidity and the temperature were measured in parallel to the audio recordings every 15 minutes, using dedicated weather loggers (Hobo, Onset, MX2302 and MX2202, Bourne, MA, USA).

### b. *Experimental protocol*

We measured sound attenuation via propagation experiments using the level difference technique (Ellinger & Hödl, 2003). This technique, which does not require the computation of free-field propagation and supports the use of a non-calibrated source and receiver, consists in measuring the difference in the frequency *spectrum* recorded at different propagation distances. The protocol first involved recording the ambient sound for 22 seconds. This recording was directly followed by the broadcasting of *white noise* for an additional 22 seconds with a wireless and waterproof loudspeaker (JBL Xtrem2, Northridge, CA, USA). The white noise was recorded at different distances by an autonomous recording unit (Song Meter SM4, Wildlife Acoustics, Maynard, MA, USA). The loudspeaker and the recorder were placed face to face at the same height (1.5 m in French Guiana, 2.5 m in Jura). The loudspeaker was calibrated to control the sound level and to flatten the broadcast signal around the desired frequency range (see Appendix B for the complete calibration procedure). White noise was broadcast at 83 dB SPL re20 µPa at 1 m in French Guiana and 78 dB SPL re20 µPa at 1 m in Jura every 10 m along a 100 m linear transect (see Appendix A for an illustration of each transect). These source levels were within the range of those previously used for other sound attenuation studies (Darras et al., 2016; Morton, 1975; Shaw et al., 2021) and corresponded to the average sound level of bird vocalizations (Aubin & Mathevon, 2020; Brackenbury, 1979). The ambient sound and the broadcast white noise were also recorded and their sound level measured using a *sound level meter* with a flat *frequency response* between 0.05 kHz and 20 kHz (SVANTEK 977A, Warsaw, Poland) positioned right next to the ARU, both microphones separated by less than 10 cm but at the same distance to the loudspeaker.

The main characteristics of the experiments are summarized in Table 1.

### c. *Sound attenuation contributions in the natural environment*

We reduced the complexity of sound attenuation in the natural environment by decomposing it into three components: (1) *geometric attenuation* ($A_{geo}$), also known as **spreading loss** or **spherical attenuation**; (2) atmospheric attenuation ($A_{atm}$); and (3) **habitat attenuation** ($A_{hab}$) (Embleton, 1996). The new concept of habitat attenuation is detailed below (see appendix C for the calculation of $A_{geo}$ and $A_{atm}$).

i. <u>Habitat attenuation</u>

The attenuation due to the habitat $A_{hab}$ encompasses the attenuation due to the **ground effect** and wave **scattering** and **absorption** by tree trunks, branches and leaves (Rossing, 2007). We did not account for the effect of topography because the transects were on flat terrain (see Appendix A). We also neglected the effect of wind as wind speed was below 5 m/s and other minor phenomena like temperature gradients and vortices as the experiments took place in closed habitats (Rossing, 2007).

We regarded the habitat as a homogeneous propagation medium governed by a single attenuation parameter which depends on the propagation distance $r$, the reference distance $r_0$ and frequency $f$. We used an empirical model of sound attenuation following an exponential decay law with a power-type frequency dependence (Szabo, 1994) whose exponent varies between 0 and 2. We assumed an exponent equal to 1, that is, a linear dependence between frequency and distance. Eventually, the habitat attenuation law is governed by a single parameter and $A_{hab}$ can be expressed in terms of sound level using the following equation, with $a_0$ being the **habitat attenuation coefficient** in dB/kHz/m:

$$A_{hab} = 20 * \log_{10}(\exp(\hat{a}_0 * f * (r - r_0)))$$
$$A_{hab} = 8.69 * \hat{a}_0 * f * (r - r_0)$$
$$A_{hab} = a_0 * f * (r - r_0) \quad (1)$$

ii. <u>Excess attenuation and experimental determination of habitat attenuation</u>

We calculated the excess attenuation $EA$ in order to estimate the difference between the theoretical attenuation and the measured attenuation through the habitat (Ellinger & Hödl, 2003). $EA$ at a given distance $r$ relative to a reference distance $r_0$ corresponds to the difference between the theoretical sound level $L$ and the experimentally measured sound level $L_{exp}$ for each frequency step $f$ and each distance step $r$, without the contribution of the ambient sound $Ln$, hereafter called $L_{exp\_corr}$ (see appendix C for the estimation of $L_{exp}$, $Ln$ and $L_{exp\_corr}$). The theoretical sound level $L$ is also equal to the initial sound level $L_0$ measured at distance $r_0$, minus the geometric ($A_{geo}$) and atmospheric ($A_{atm}$) attenuations. Thus, $EA$ can be expressed as:

$$EA = L_0 - A_{geo} - A_{atm} - L_{exp\_corr} \quad (2)$$

If we consider that the excess attenuation $EA$ is entirely due to the habitat attenuation $A_{hab}$, then $EA = A_{hab}$. We obtain the expected frequency linear dependency of $A_{hab}$ after normalizing $EA$ by the propagation distance $r - r_0$ for each distance $r$:

$$EA = a_0 * f * (r - r_0)$$
$$EA / (r - r_0) = a_0 * f \quad (3)$$

We estimated $a_0$ by carrying out a linear regression using a robust (i.e. minimizing the contribution of outliers) linear model of Eq. 3 as a function of $f$, $r$ and $r_0$. The reference distance $r_0$ was set to vary between 10 m and 50 m (over 50 m, the signal is too weak to be used as a reference signal) while the range of the distance $r$ was set between 10 m to 100 m. We did not use distances smaller than 10 m because significant distortion due to the ground effect could occur (Embleton, 1996). We restricted the frequency range of the linear regression between 0.5 kHz and 15 kHz, because outside this range most of the propagated white noise $L_{exp}$ was not separable from the ambient sound $Ln$. The estimate was validated if the linear model was highly significant (p < 0.001). We obtained $a_0$ for both habitats with the ARU and the sound level meter. The final value $a_0$ corresponded to the average of the $a_0$ resulting from the linear minimizations carried out by varying the reference distance $r_0$ between 10 m and 50 m.

   iii. <u>Full propagation model</u>

To describe and predict the attenuation of acoustic waves propagating through both forests, we simultaneously took into account the three types of attenuations described above. We used the following complete propagation model:

$$L = L_0 - A_{geo} - A_{atm} - A_{hab} \qquad (4)$$

In order to validate the full attenuation model, theoretical propagation curves were obtained by applying the Eq. 4 for four frequency bands (i.e. [0−5] kHz, [5−10] kHz, [10−15] kHz and [15−20] kHz) for each habitat and each recording device. $L_0$, $A_{geo}$, $A_{atm}$ and $A_{hab}$ were calculated using the experimental parameters (i.e. temperature, relative humidity, barometric pressure, initial sound level) and the estimated coefficients $a_0$ (see appendix C for calculation of $A_{geo}$ and $A_{atm}$). Finally, the ambient sound level $Ln$ was taken into account by adjusting its contribution to each theoretical propagation curve $L$ in order to estimate the experimental sound level $L_{exp}$ which is a combination of the propagated sound (i.e. the white noise signal) and the ambient sound that arrive simultaneously at the microphone. The result was then compared to the sound levels $L_{exp}$ measured for each propagation distance and each frequency band. In the case of ARU, we applied a frequency correction in order to flatten as much as possible the frequency response of the ARU (see appendix D for the procedure of calibration). Finally, for a demonstration purpose, we overlaid two curves obtained from partial attenuation model (1) with only the geometric attenuation (i.e. $L_0 - A_{geo}$) and (2) with both geometric and atmospheric attenuations (i.e. $L_0 - A_{geo} - A_{atm}$).

   d. *Detection distance estimation by autonomous recording units*

We adopted the following definition of the detection distance: a sound emitted at a given frequency is considered detectable but not necessarily discriminable if its sound level is greater than a threshold just above the average sound level of the ambient sound corresponding to that frequency (Dawson & Efford, 2009). We inferred the ARU detection distance by finding the distance $r$ that minimizes the difference between the propagated sound level $L$ (see Eq. 4) and the level of the ambient sound $Ln$ (Ellinger & Hödl, 2003).

In the case of data collected during a propagation distance experiment, $Ln$ corresponds to the average sound level measured during the ambient sound recording before broadcasting the white noise signal. In the case of the annual data sets, we set the ambient sound as the median of the sound pressure level measured at each 1 kHz frequency bin within the frequency range 1–20 kHz, for each month within a 1-hour slot corresponding to four 1-minute-long recordings. By knowing the ambient temperature and the relative humidity measured by the loggers and using the propagation model defined in Eq. 4, we were able to calculate the detection distance of 80 dB SPL of white noise and focused on the frequency range 1–8 kHz, where most of the biophony occurred. We made the assumption that the coefficient of attenuation $a_0$ of each habitat remained constant throughout the year as there were only minimal changes in habitat structure over the course of the year.

   e. *Software tools*

Data processing and statistical analyses were performed in R 3.6.3 (TEAM, R.Core, 2020) with the seewave (Sueur et al., 2008) and tuneR (Ligges, 2016) packages for signal processing, the robustbase package for linear regression (Maechler et al., 2022) and the plotly (https://plotly.com/graphing-libraries/) and tidyverse (Wickham et al., 2019) packages for graphs.

The ambient sound estimation over the course of a year in a Neotropical rainforest and an Alpine coniferous forest was performed in Python with the scikit-maad package (Ulloa et al., 2021).

Comprehensive functions to compute sound pressure level from audio files, spreading loss, atmospheric attenuation and habitat attenuation as well as to estimate the detection distance are available on GitHub (https://github.com/shaupert/HAUPERT_MEE_2022) (R) and in the scikit-maad package (Python).

## 3. Results

### *a. Ambient sound and white noise sound level variation as a function of distance*

The time-frequency representation of the sound recorded by the ARU (see Appendix E for results obtained with the sound level meter) as a function of distance revealed that in both forests: 1) the frequency spectrum of the white noise signal recorded by the ARU was not flat; 2) part of the white noise signal disappeared in the ambient sound at 30 m and beyond (Figs. 1a and 2a).

Similarly, the mean power spectra of the white noise showed ***aperiodic lobes*** that probably resulted from a ***comb filter*** effect which delayed version of the transmitted signal to itself, causing constructive and destructive ***interference*** due to the reverberation of the sound on the ground (Figs. 1b and 2b).

The overall sound level (ambient + white noise) as well as the ambient sound level alone and the white noise level alone in the Neotropical forest revealed a peak of acoustic energy of between 6.5 kHz and 8.5 kHz, followed by a smaller peak of around 15 kHz, due to insect stridulations. Likewise, a peak of acoustic energy of between 0 and 1 kHz occurred in the Alpine coniferous forest, mostly due to aircrafts and the rustling of leaves (Figs. 1c and 2c).

In both forests, the sound level of the white noise decreased as expected with distance to reach a value close to the ambient sound level beyond 60 m and for frequencies above 10 kHz.

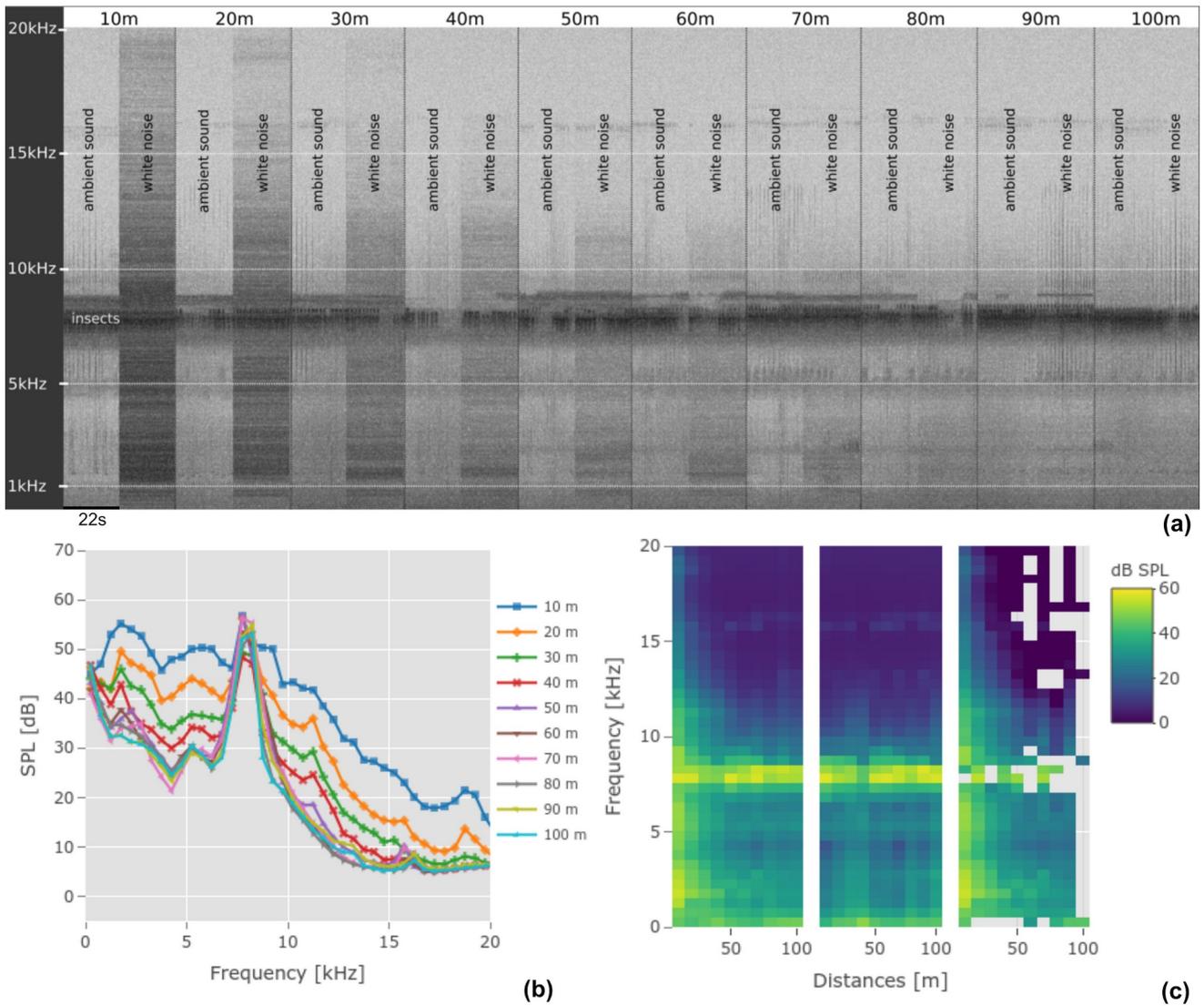

*Figure 1: Ambient sound and white noise variations according to distance in the Neotropical rainforest (French Guiana). (a) Time-frequency representation of the sounds recorded by the ARU of the ambient sound and the white noise between 10 m and 100 m. (b) Mean spectra of the total sound pressure level (dB SPL) (i.e. ambient sound + white noise) measured at each distance. (c) Map of sound levels: overall transmitted signal at each propagation distance (left), ambient sound (centre) and white noise (right). The missing pixels in panel (c) represent the distance-frequency combinations where the amplitude level of white noise could not be disentangled from the ambient sound. The unit is dB SPL (re20 µPa). The audio signals were collected by the ARU on 18 February 2019 between 2 pm and 5 pm.*

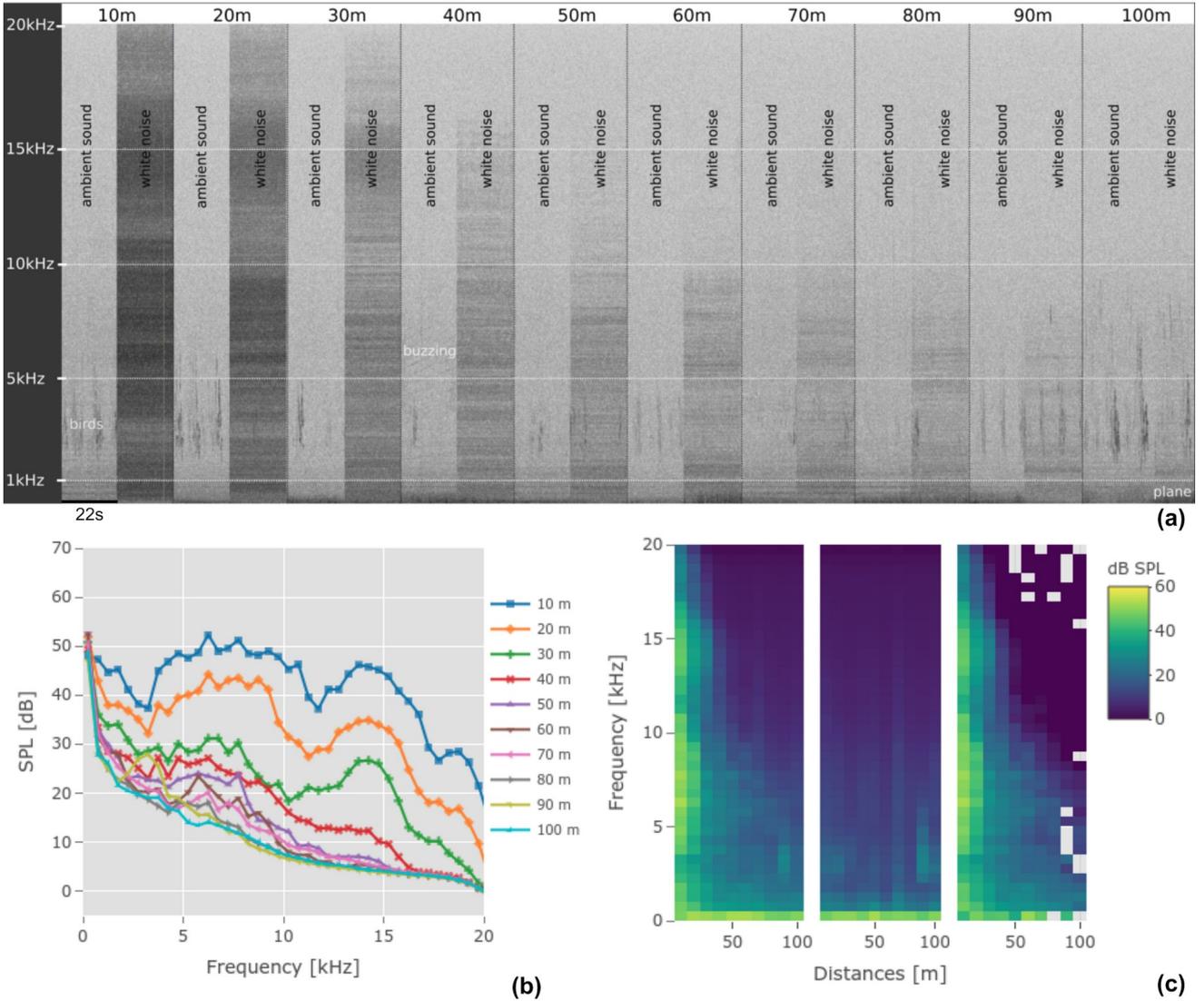

*Figure 2: Ambient sound and white noise variations according to distance in the Alpine coniferous forest (Jura). (a) Time-frequency representation of the sounds recorded by the ARU of the ambient sound and the white noise between 10 m and 100 m. (b) Mean spectra of the total sound pressure level (dB SPL) (i.e. ambient sound + white noise) measured at each distance. (c) Map of sound levels: overall transmitted signal at each propagation distance (left), ambient sound (centre) and white noise (right). The missing pixels in panel (c) represent the distance-frequency combinations where the amplitude level of white noise could not be disentangled from the ambient sound. The unit is dB SPL (re20 µPa). The audio signals were collected by the ARU on 11 July 2019 between 11 am and 1 pm.*

### b. Habitat attenuation

The linear dependence between the habitat attenuation $A_{hab}$ and the frequency $f$ was verified using the excess attenuation $EA$ normalized by the propagation distance ($EA / (r – r_0)$) for different distances of reference $r_0$ (see Eq. 3). Moreover, the linear dependence between frequency and ($EA / (r – r_0)$) only occurred when the ambient sound was subtracted from the original sound level, while this was not true when no correction was applied (see appendix G).

The habitat attenuation coefficients $a_0$ are summarized in Table 2. The average value of the attenuation coefficient $a_0$ in the Neotropical rainforest was 0.019 ± 0.001 dB/kHz/m according to the sound level meter and 0.011 ± 0.001 dB/kHz/m according to the ARU. In the Alpine coniferous forest, the attenuation coefficient $a_0$ was 0.020 ± 0.008 dB/kHz/m according to the sound level meter and 0.024 ± 0.008 dB/kHz/m according to the ARU. The dispersion of the values of $a_0$ around the mean value precluded any statistical analyses. However, attenuation was slightly lower in the Neotropical rainforest (French Guiana) than in the Alpine coniferous forest (Jura).

The theoretical propagation curves in the Neotropical rainforest using the full attenuation model (i.e. $A_{geo}+A_{atm}+A_{hab}$) and partial attenuation models (i.e. $A_{geo}$ only or $A_{geo}+A_{atm}$) are shown in Fig. 3 when the receiver is the ARU (see Appendix F for the results obtained in the case of the theoretical propagation curves in the Alpine coniferous forest and when the receiver is the reference sound level meter). The trend of the sound level attenuation predicted by the full attenuation model, including the addition of the ambient sound, was in good agreement with the experimental values, except for the highest bandwidth 15–20 kHz, where the model overestimated the experimental values by 6 dB (see Fig. 3). The discrepancy between the theoretical propagation curves and the experiment in the case of the ARU might have been due to the loss of sensitivity of the microphone in high frequencies.

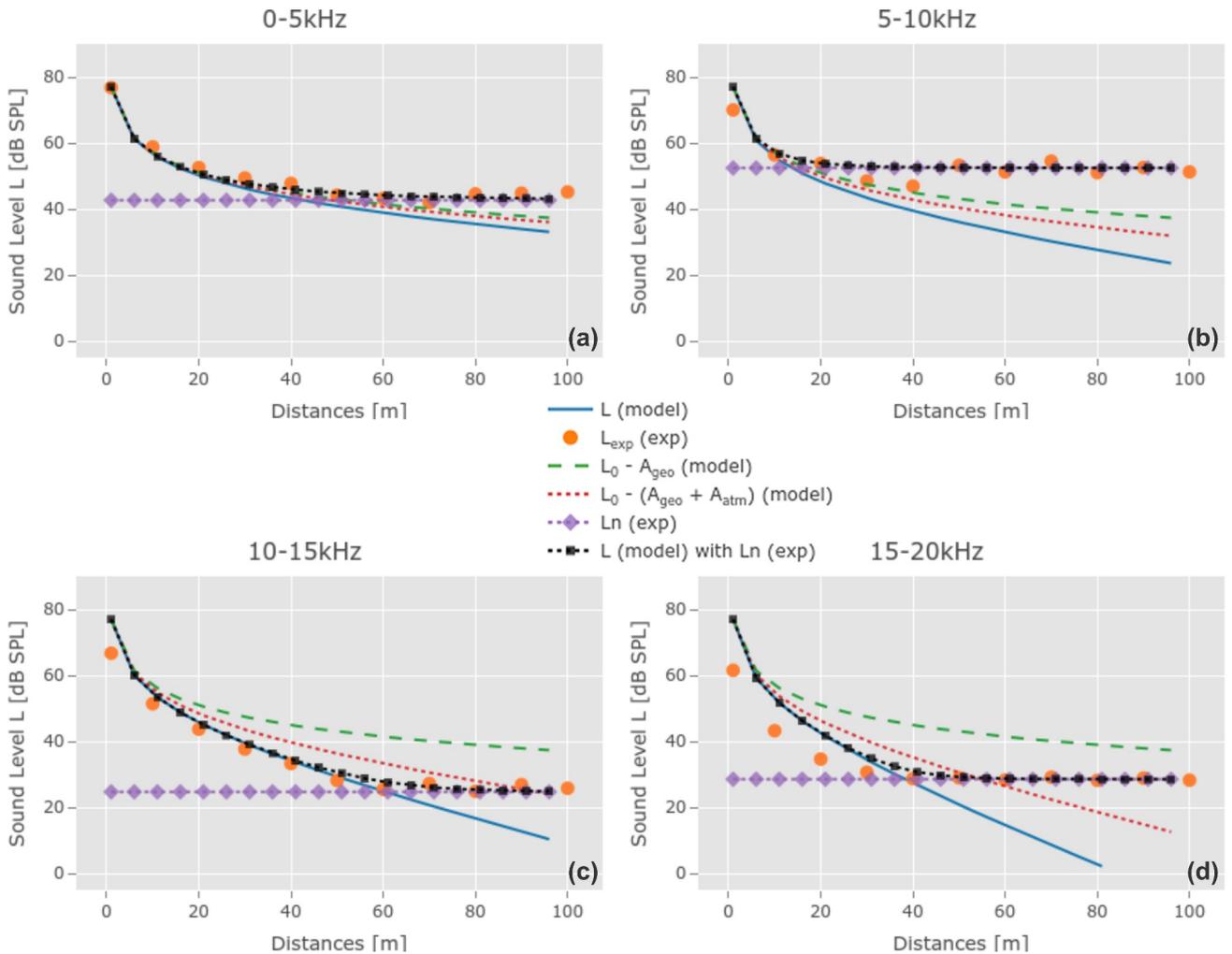

*Figure 3: Comparison between experimental data acquired with the ARU in a Neotropical rainforest (French Guiana) and theoretical propagation curves obtained from the full model as described by Eq. 4. The comparison was performed for the following bandwidths: (a) 0–5 kHz; (b) 5–10 kHz; (c) 10–15 kHz; and (d) 15–20 kHz. The habitat coefficient attenuation parameter was $a_0 = 0.011$ dB/kHz/m. L (model): predicted sound level according to the full propagation model; $L_{exp}$ (exp): experimental sound level; $L_0$ - $A_{geo}$ (model): predicted sound level according to a partial propagation model taking into account only the geometric attenuation; $L_0$ - ($A_{geo}$ + $A_{atm}$) (model): predicted sound level according to a partial propagation model taking into account both, the geometric and atmospheric attenuation; $L_n$ (exp): experimental ambient sound level; L (model) with $L_n$ (exp): predicted sound level according to the full propagation model taking into account the contribution of the experimental ambient sound level*

### c. Detection distance estimation by autonomous recording units

The detection distance was driven by the combination of multiple factors: the source sound level $L_0$, the source frequency bandwidth, the sound attenuation factors (geometric $A_{geo}$, atmospheric $A_{atm}$ and habitat $A_{hab}$ attenuations), and the ambient sound level $L_n$.

#### i. Contributions of the different attenuation factors

Regardless of the type of habitat, the most important attenuation factor was $A_{geo}$ (Fig. 4). Its contribution was predominant for low frequencies, especially in the Neotropical rainforest, where $A_{geo}$ was the main sound attenuation factor, representing more than 75 % of the attenuation for frequencies below 10 kHz. For frequencies above 10 kHz, the relative proportion of $A_{geo}$ decreased with frequency to become almost of the same order of magnitude as the sum of the other two types of attenuation, $A_{atm}$ and $A_{hab}$, especially in the Alpine coniferous forest. The contribution of atmospheric attenuation $A_{atm}$ followed the opposite trend to $A_{geo}$ as it increased with frequency. This trend was similar for both forests. The contribution of attenuation due to the habitat $A_{hab}$ was of the same order of magnitude as $A_{atm}$, in particular in the Alpine coniferous forest for frequencies above 10 kHz. The contribution of $A_{hab}$ for frequencies below 10 kHz was slightly smaller in the Neotropical rainforest than in the Alpine coniferous forest.

#### ii. Variation in detection distance according to frequency

The detection distance of white noise at 80 dB SPL re20 µPa varied considerably depending on the frequency and the type of habitat. In the Neotropical rainforest, the detection distance varied between a minimum of 3 m at 7.75 kHz and a maximum of 83 m at 4.25 kHz (Fig. 4a). At a large frequency scale, the average detection distance was 44 m and 45 m for the frequency range 0–10 kHz and 10–20 kHz, respectively. In the case of the Alpine coniferous forest, the detection distance varied between a minimum of 6 m at 0.25 kHz and a maximum of 125 m at 1.75 kHz (Fig. 4b). The detection distance was on average 81 m for the 0–10 kHz frequency band and almost three times shorter with an average of 35 m for the 10–20 kHz frequency band. With regard to estimating the detection distance by frequency of the quietest and the loudest passerine birds in Jura by the ARU, see Appendix H.

#### iii. Variation in detection distance according to time of day and season

The detection distance within both habitats varied according to the diel period, exhibiting a clear difference between night and day (Figs. 5 and 6). In the case of the Neotropical rainforest, the detection distance varied considerably during the night and day cycle and by season. The detection distance was lowest – just a few metres – for the frequency range 7–8 kHz occupied by insects. In the case of the Alpine coniferous forest, the detection distance varied most for frequencies below 2 kHz and least for frequencies above 4 kHz. The detection distance fluctuated by up to five times according to the season and the time of the day.

iv. <u>Comparison between ARU and reference sound level meter</u>

In both forests, the detection distance estimated by the ARU was similar to that estimated by the reference sound level meter (see Appendix H). The detection distance was slightly lower for the ARU than for the sound level meter, in particular for frequencies above 10 kHz. This behaviour owed to the frequency response of the ARU which was not flat and which decreased significantly above 10 kHz (see Appendix C). This resulted in a lower SNR for high frequencies, reducing the sensitivity threshold of sound attenuated by a long propagation distance.

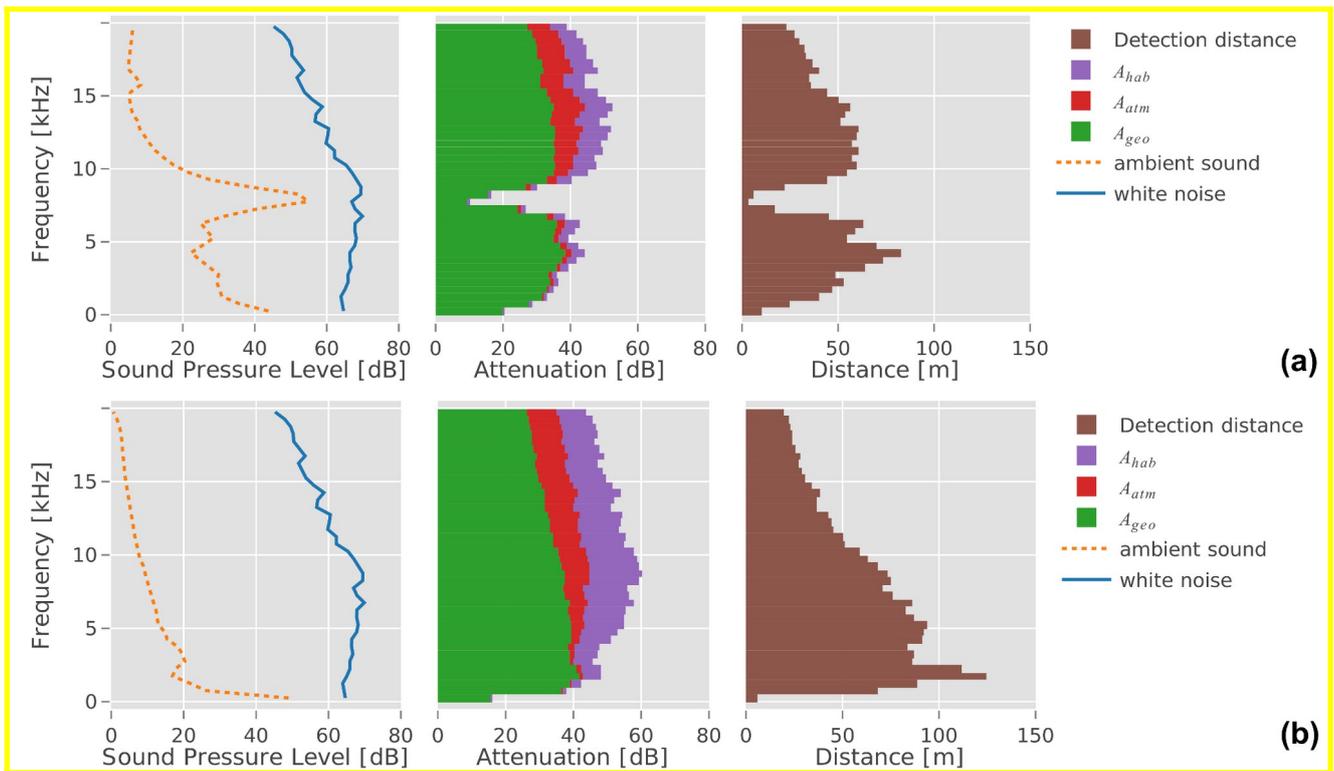

*Figure 4: Detection distance of the ARU obtained with the model in: (a) a Neotropical rainforest (French Guiana); (b) an Alpine coniferous forest (Jura, France). The source was a wideband (0–20 kHz) white noise broadcast at 80 dB SPL re20 μPa at 1 m. The profile of the source and the ambient sound takes into account the frequency response of the ARU. The contribution of geometric (i.e. spreading loss) attenuation ($A_{geo}$, green), atmospheric absorption ($A_{atm}$, red) and habitat attenuation ($A_{hab}$, purple) are depicted with different colours.*

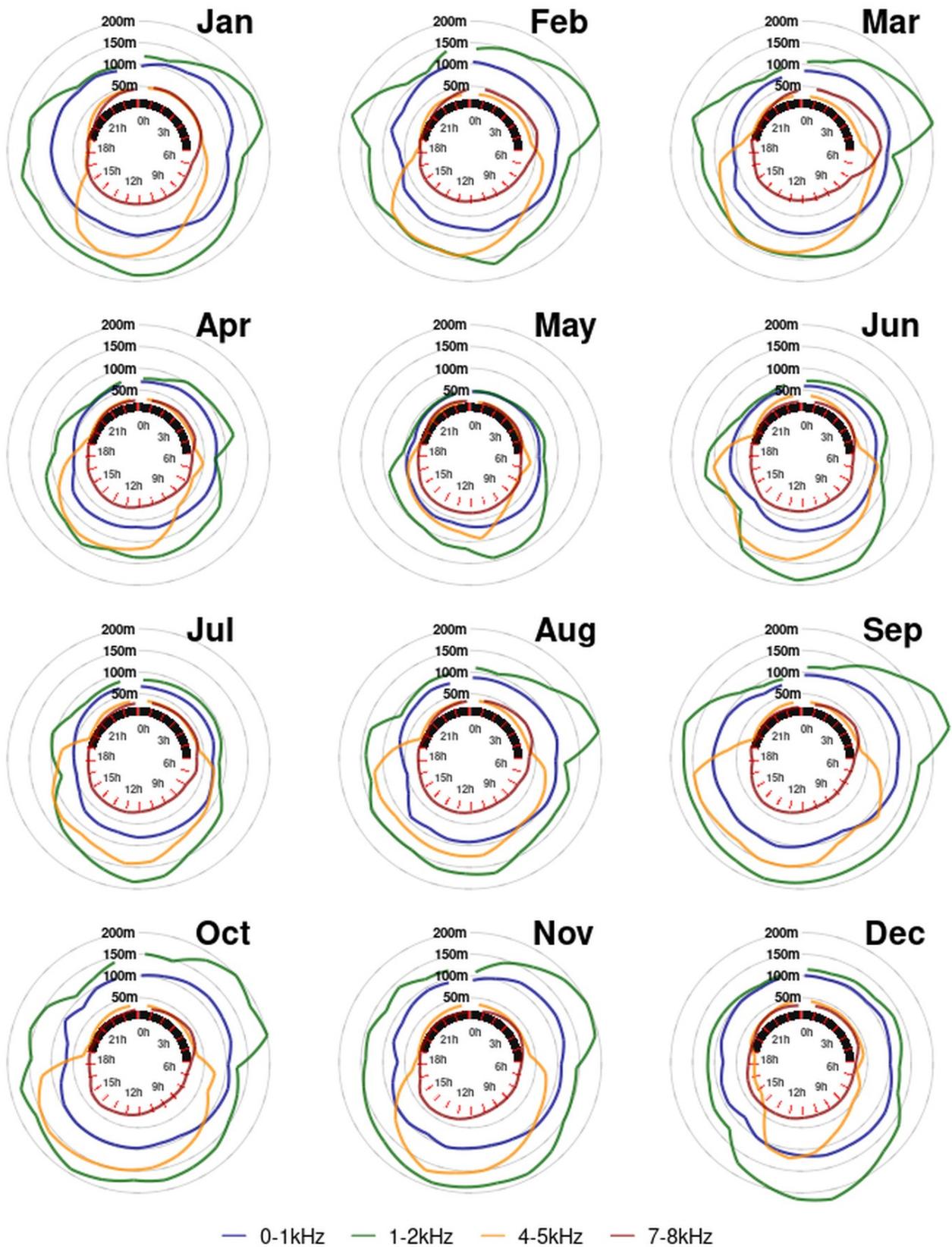

*Figure 5: Detection distance variation of the ARU according to night and day cycle and month in a Neotropical rainforest (French Guiana) taken into account the variation of the environmental factors (T °C and RH %) as well as the ambient sound level Ln. The sound source was a wideband (0–20 kHz) white noise broadcast at 80 dB SPL re20 µPa at 1 m. Each circular plot represents one month, with each circle corresponding to a propagation distance from 50 to 200 m by a step of*

50 m over a complete day (white) and night (black) cycle. Line colours refer to four 1 kHz frequency bands (0–1 kHz, 1–2 kHz, 4–5 kHz, 7–8 kHz).

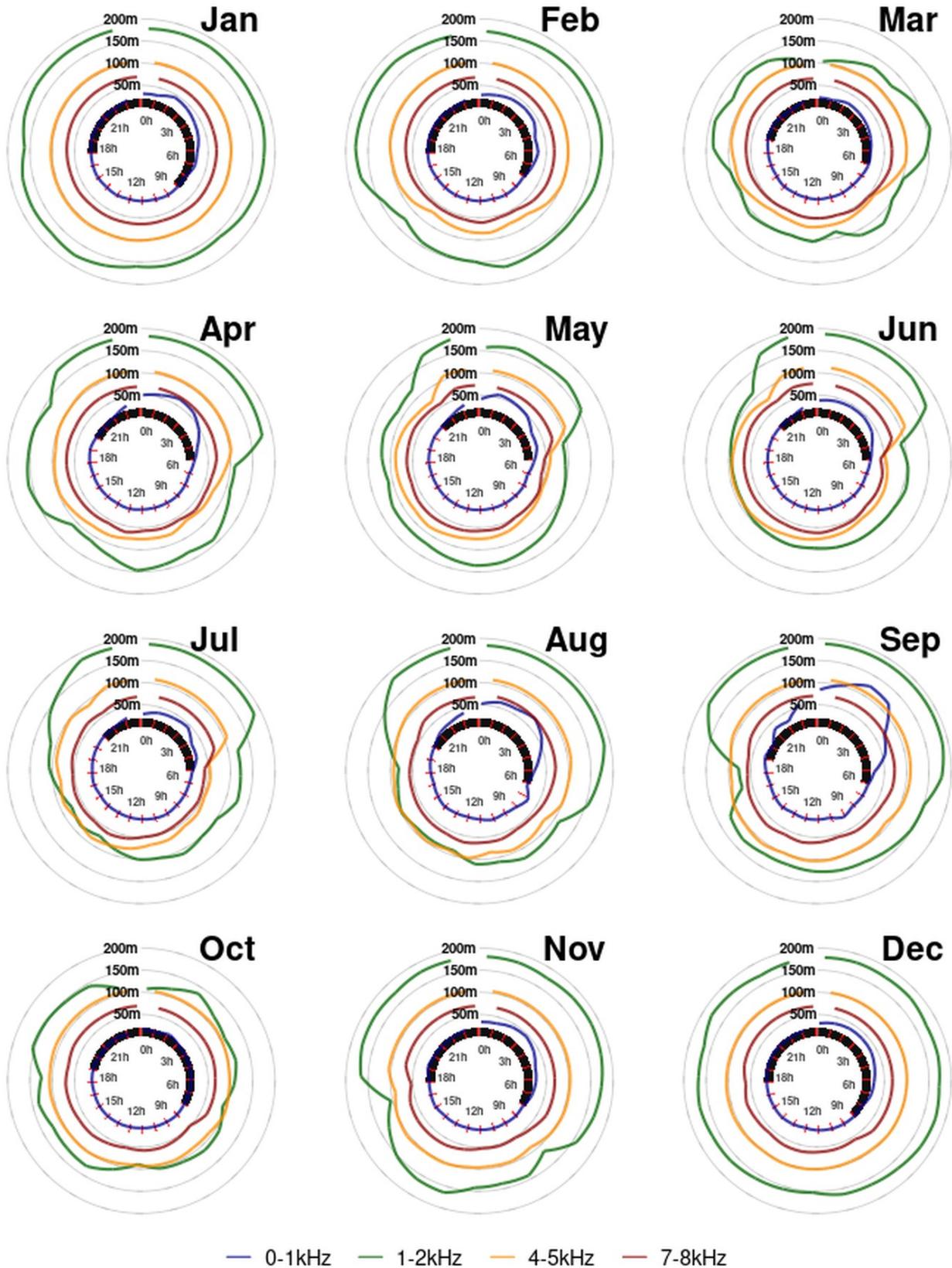

*Figure 6: Detection distance variation of the ARU according to night and day cycle and month in an Alpine coniferous forest (Jura) taken into account the variation of the environmental factors (T °C and RH %) as well as the ambient sound*

*level Ln. The sound source was a wideband (0–20 kHz) white noise broadcast at 80 dB SPL re20 µPa at 1 m. Each circular plot represents one month, with each circle corresponding to a propagation distance from 50 to 200 m by a step of 50 m over a complete day (white) and night (black) cycle. Line colours refer to four 1 kHz frequency bands (0–1 kHz, 1–2 kHz, 4–5 kHz, 7–8 kHz).*

## 4. Discussion

In this study, we propose a standardized and repeatable method based on the excess attenuation framework to estimate the detection distance of an autonomous recording unit (ARU). After taking into account the ambient sound and the atmospheric absorption frequency dependence, in two types of forests, we show that the attenuation of a habitat can be modelled with a single parameter $a_0$ expressed in dB/kHz/m. We also confirm that for a given habitat, the detection distance varies depending on the initial sound pressure level of the broadcast signal and its frequency bandwidth. Moreover, we highlight that the main factor that alters the detection distance is the masking effect of the ambient sound, which may vary dramatically during the 24-hour cycle and across seasons in both habitats. These findings are significant in providing reliable biodiversity measures regardless of whether acoustic surveys are performed with ARUs or by traditional human point count surveys.

### a. *Determining sound attenuation requires consideration of ambient sound and atmospheric absorption frequency dependence*

We first demonstrate that besides geometric attenuation ($A_{geo}$), it is necessary to take into account atmospheric absorption ($A_{atm}$) in order to compute the excess attenuation *EA* and be able to estimate the habitat attenuation ($A_{hab}$). We show that this corresponds to about a fifth of $A_{geo}$ for a mid-frequency range relative to bird calls and songs and up to a third for frequencies higher than 10 kHz.

We also confirm as has already been noted (Darras et al., 2016; Shaw et al., 2021) that it is necessary to remove the contribution of the ambient sound from the sound propagated through the habitat to estimate without bias the attenuation process that affects sound propagation.

Altogether, we strongly recommend removing the contribution of the ambient sound *Ln* as well as subtracting the attenuation $A_{atm}$ and $A_{geo}$ relative to the distance and the frequency band, in order to extract the coefficient of attenuation of the habitat $a_0$. If not, *EA* exhibits no linear relationship between frequency and distance (see Figs. G-2 and G-3 in appendix G).

### b. *Sound attenuation partially depends on the habitat and can be modelled by a single value $a_0$*

Our propagation measurements reveal that sound attenuation partially depends on the habitat following an exponential decay that can be modelled by a single coefficient of attenuation, $a_0$. To our knowledge, this is the first sound attenuation model of natural environments that has been validated over a large frequency band (0–20 kHz), fitting particularly well in the 1–10 kHz frequency bandwidth, where most of the biophony is found. Our model, which is based on acoustic physics and not descriptive statistics, does not attempt to evaluate separately the different sources of attenuation occurring within the habitat, but rather quantify the role of each variable, including temperature, relative humidity, atmospheric pressure, distance, frequency, ambient sound pressure level and habitat.

Having a single parameter opens the possibility of comparing attenuation among different habitats and tracking the variation in habitat attenuation over seasons in case of deciduous forests, or over decades in case of vegetation's alterations (e.g. increase/decrease of the density of trees, timber extraction). We found that the attenuation coefficient $a_0$ did not differ significantly between the Neotropical rainforest and the Alpine coniferous forest, a counter-intuitive result that is actually in agreement with the literature (Aylor, 1972; Ellinger & Hödl, 2003; Price et al., 1988). The apparent similarity of $a_0$ between both forests might be explained first by the measurement uncertainty possibly owing to an imperfect alignment between the loudspeaker and the microphone. The apparent similarity of $a_0$ might also be explained by the relatively equal influence of the three main attenuation factors expected to play a role in sound propagation: (1) the ground effect, which acts as a comb filter producing typical frequency bumps and rays, is mostly visible at the lowest frequencies (<1 kHz) (Tarrero et al., 2008), minimizing its influence on the overall attenuation law; (2) the vegetation effect, which is expected to induce scattering at high frequencies (>1 kHz), actually seems to play a similar role in sound degradation regardless of forest type (Aylor, 1972; Darras et al., 2016; Shaw et al., 2021); and (3) the micro-meteorological effect, including turbulence, which is stronger in open areas than in closed areas like the forest understorey (Wiener & Keast, 1959).

Due to the similarity of $a_0$ for two different type of forests, we make the assumption supported by the results found in the literature (see Appendix G) that it is possible to use an average value of $a_0$ (i.e. 0.018 dB/kHz/m) as a reasonable approximation to model part of the attenuation of sound through forests with relatively sparse foliage (i.e. with a visibility around 10 m), regardless of the type of forest. For instance, this opens the possibility of obtaining a first approximation of the optimal spatial sampling when designing new passive acoustic surveys. In temperate forests such Alpine coniferous forests, a minimum grid spacing of 500 m (i.e. a 250 m radius) between ARUs is recommended when the louder bird vocalization in the habitat reaches up to 90 dB SPL (see appendix H). We advise researchers to estimate the attenuation coefficient of their habitat of interest using this method in order to verify this assumption and adjust $a_0$ to their need.

c. *Drivers of the detection distance*

We show that it is possible to estimate the distance at which a transmitted sound vanishes. We call this distance the detection distance, in contrast to the active distance, which is from the point of view of the emitter (i.e. the source). The detection distance also differs from the sampling distance (also referred to as the detection radius or the detectability area) used in human or automatic point count surveys (MacLaren et al., 2018; Pérez Granados et al., 2019; Yip et al., 2017; Yip & Bayne, 2017). Indeed, in this study we do not provide the probability of detecting a vocalization at a relative distance from the receiver, but rather provide the distance at which the vocalization no longer emerges above the ambient sound. Despite this conceptual difference, the detection distance found in both habitats is consistent with the sampling distances found in the literature (Darras et al., 2018; MacLaren et al., 2018; Shaw et al., 2021).

For a given natural and relatively homogeneous environment over the period of observation, that is, with a fixed $a_0$, we show that the detection distance mostly depends on the ambient sound. A clear example are the insect stridulations found between 6.5 kHz and 8.5 kHz in the Neotropical rainforest, which are so strong that the detection distance in this frequency range is just a few meters away from the receiver. Environmental factors such as temperature and relative humidity play only a minor role in the global sound attenuation process through the variation of the atmospheric absorption (see appendix J to observe the variation of the detection distance over the diel cycle and across seasons when the ambient sound level $Ln$ is fixed to the average yearly value for each frequency bins and hours of the day). Moreover, we demonstrate for the first time that the detection distance may vary considerably during the night and day cycle and by month regardless of the habitat and the nature of the ambient sound. For instance, the detection distance within a 24-hour period was never close to a constant in French Guiana, whatever the month, its patterns being more complex than the day-night alternation. In Jura, the detection distance for the frequency bin 0–1 kHz increased by a factor of five between the day and the night in relation to aircraft traffic (Grinfeder et al., 2022). In both forests, the detection distance varied by up to a factor of 5 depending on the frequency bandwidth. The detection area, which is the square of the distance (i.e. the radius), might then vary by up to a factor of 25 depending on the hour of the day and the season. This major result implies carefully evaluating the detection distance at the frequency level in order to provide reliable information on species density when using ARUs or human point count methods. Variability among surveys with respect to the ambient sound might also affect most distance sampling or detection probability models, resulting in an over- or underestimation of species abundance or species richness (Koper et al., 2016). Although the negative effect of the natural ambient sound on avian detectability is well-known for point counts (Simons et al., 2007) and ARU surveys (Darras et al., 2016), to our knowledge few previous studies have explicitly included the ambient sound as a variable in their models to predict detectability (Anderson et al., 2021) and none of them have shown that it can actually be generalized to other ambient sound levels at different frequency ranges. We advise measuring the ambient sound by frequency bands in parallel to ARUs or point count surveys in order to be able to take into account the variation in the model according to the frequency.

In addition to the ambient sound, the detection distance depends on the sound level and the frequency content of the source. In this study, we did not test the role of the sound level, as we used a fixed wideband sound source level. Nevertheless, we are aware that the sound levels of biological sources, although rarely known, vary across species, among individuals of the same species, and within the same individual. For instance, the sound level of common birds can vary from 75 dB SPL at 1 m in the case of the goldcrest (*Regulus regulus*) (Brackenbury, 1979), among the quietest bird in Jura, to up to 111.5 dB SPL at 1 m in the case of the screaming piha (*Lipaugus vociferans*) (Nemeth, 2004), among the loudest bird in French Guiana. In this study, we instead focused on the frequency dependence of the detection distance. We have shown that in both habitats, the detection distance of high-frequency sound is generally shorter than mid-frequency sound. This is congruent with a signal processing perspective, which states that a habitat acts as a low-pass filter (Römer, 2001), as higher frequencies with smaller wavelengths suffer more from multiple scattering than do lower frequencies. Interestingly, in the case of low frequency sound (< 1 kHz), the detection distance is smaller than in the case of mid frequency sound. The detection distance of low-frequency sound is mostly driven by the ambient sound, which is louder than at a mid- or high-frequency range, rather than by the habitat characteristics. The main origin of low-frequency ambient sound may be associated with geophonic sounds (e.g. wind or distant rain), anthropogenic sounds (e.g. aircrafts or cars) or technical issues (e.g. recorder noise).

Finally, we have shown that the frequency response of ARUs, which is rarely flat, is also a driver of the detection distance. For instance, the frequency response of the SM4 ARU decreases steeply above 10 kHz, limiting the detection distance of high-frequency content sources compared to devices with a flat frequency response, such as sound level meters. This result is supported by previous studies comparing different ARUs (Darras et al., 2016; Pérez Granados et al., 2019; Rempel et al., 2013). The quality of the electronics, especially the flatness of the ARU's microphone, should be carefully considered when a correct detection distance is required across a large frequency bandwidth (Turgeon et al., 2017).

### d. *Limitations of the study*

Our experiment was based on a single transect facing the microphone. In this configuration, we did not have access to the propagation distance in other directions from the recorder. However, the direct sound path between the loudspeaker and the ARU might have resulted in the longest possible detection radius range, which is required to determine the minimum distance between ARUs to avoid violating the assumption of independent observations. Furthermore, focusing on a single transect seems to be a reasonable option for homogeneous habitats and when several ARUs are deployed and extensive propagation experiments cannot be run around each of them.

In addition, we did not take into account the directivity of the microphone, especially given that the recorders were mounted on tree trunks, which can cause sound shadows (Darras et al., 2018). The detection distance was probably neither circular nor planar due to the specification of the microphone, the position and orientation of the ARU and the landscape (Castro et al., 2019; MacLaren et al., 2018; Shaw et al., 2021). However, as our model directly predict the attenuation of sound depending on the physical laws that explicitly govern the sound propagation in natural environments, this opens the possibility to include, in a second step, additional attenuation factors such as the angle (0° to 360°) between the source and the microphone. Our model may need to be adapted to mountainous or urban areas – where the trajectory of the sound is more difficult to predict due to reverberation – and to open areas, where winds can dominate. Moreover, our model does not take into account the fact that sound is propagated in three dimensions, hence different attenuation laws may also occur in the vertical direction (Ellinger & Hödl, 2003). For instance, it has been observed that blackbirds *(Turdus merula)* choose to sing at higher elevation spots to limit the degradation of their songs (Dabelsteen et al., 1993).

From an ecological statistics perspective, our physics-based model alone is not sufficient to provide a reliable probability of detection of individuals, due to the variability of the sound sources directivity. Indeed, the estimation of the detection distance might be overestimated when the individuals do not vocalize towards the microphone which might lead to an underestimation of the number of individuals within a fixed radius. We believe that this bias might be overcomed by developing a dedicated statistical model of detection probability that could be partially based on our physics-based model.

Finally, we studied signal detection based on a threshold derived solely from the amplitude of the sound. A sound is set to be detected if its level exceeds that of the ambient noise. Such a threshold is suitable for the automatic processing of *spectrograms* where the phase is generally neglected. However, it does not take into account more complex cues (such as repetition, duration and modulations) used by animals and humans (Aubin & Jouventin, 1998; Aubin & Mathevon, 2020) for detecting and recognizing sounds in noisy environments. Our estimation is therefore conservative, meaning that the detection distance may be longer when more complex cues are included.

## 5. Perspectives and conclusion

Passive acoustic monitoring based on the deployment of ARUs appears to be a reliable tool for assessing acoustic biodiversity over the long term and at large spatial scales in a non-invasive way (Castro et al., 2019; Darras et al., 2018, 2019; Van Wilgenburg et al., 2017; Yip et al., 2017). Recently, species distribution models (SDMs) adapted to passive acoustic surveys have been proposed based on either the acoustic space occupancy model (Rappaport et al., 2020), the modified spatial capture-recapture model (MacLaren et al., 2018) or integrated models combining acoustic and point count data (Doser et al., 2021; Van Wilgenburg et al., 2017). They all infer sound attenuation from audio recordings as covariates without explicitly describing the physical process behind the attenuation, limiting their versatility, especially when the sampling distance varies with ambient sound. Incorporating some knowledge of sound detection distance into these models, for example habitat attenuation and ambient sound, would increase the robustness of the next generation of SDMs.

The new method we propose may be extended to point count surveys performed by human experts in order to avoid, for instance, inaccurate estimations of bird distances. The ARU transfer function can be directly replaced by the average human ear's sensitivity – known as the equal-loudness contour (Fletcher & Munson, 1933) – and approximated by the A-weighting curve (Marsh, 2001). Similarly to ARUs, the human ear's performance is best around 1 khz to 5 kHz.

Throughout this article, we have referred to the detection distance. However, we can reverse this perspective and take the place of the emitter instead, so that the active distance of the emitter, that is, a vocalizing animal, can be estimated. If we accept the concept of reciprocity, the exact same model can be applied, because sound propagation is a reversible phenomenon. Within the context of intra- or inter-species communication, the ARU transfer function may be swapped with the hearing transfer function of the target species in order to simulate the intra-species active space. As an example, we simulated the attenuation of the song of the screaming piha (*Lipaugus Vociferans*) from 200 m to 700 m through the tropical rainforest. We were able to observe that the high-frequency content disappears first and that the low-frequency content travels up to 666 m (see Appendix J). Modelling such attenuation provides the possibility of tackling several questions in bioacoustics and ecoacoustics, such as the use of public and private information (Aubin & Mathevon, 2020) and the acoustic adaptation hypothesis (Morton, 1975).

Tuning the signal to noise ratio (SNR) between the foreground signal of interest (i.e. bird songs or other animal vocalizations) and the background signal (i.e. the ambient sound) is key when preparing training data sets to train models such as convolutional neural networks (CNNs) for automatic species identification (Kahl et al., 2021). Simulating the propagation attenuation of the foreground signal would improve such data augmentation and thereby provide more realistic soundscapes. In the opposite direction, very recently a CNN was trained with spectrograms of a single call recorded at different distances to infer the sampling detection of that call (Yip et al., 2020).

Finally, we think that our model opens the possibility of estimating the sound pressure level of most vocalizing species, which remain poorly documented due to difficulties in obtaining data at the reference distance (i.e. 1 m). An accurate assessment of the sound pressure level requires the estimation of the full attenuation law along the direct path between the animal and the microphone – which can be provided by our method – as well as a correct estimation of the distance between the animal and the microphone.

To summarize, we have proposed a standardized, repeatable and generalized physics-based method for determining outdoor sound attenuation in two very different kinds of forests, opening the possibility to extend this method to other terrestrial environments. We have demonstrated that the attenuation of sound in different forests can be summarized, as a first approximation, by a single value. We have shown the detection distance of an ARU can also be predicted and may significantly fluctuate because of the variation in the ambient sound. Therefore, we strongly recommend considering the variation of the detection distance over the diel cycle and across seasons in order to provide reliable and operational biodiversity indicators for conservation policy and wildlife management planning.

**References**

Anderson, E. K., Kerby-Miller, P. F., Pupko, J. M., Sharp, N. R., Tolan, K. S., & Hill, J. M. (2021). *Accounting for point count ambient noise increases population size estimates* [Preprint]. Ecology. https://doi.org/10.1101/2021.04.06.438644

Aubin, T., & Jouventin, P. (1998). Cocktail–party effect in king penguin colonies. *Proceedings of the Royal Society of London. Series B: Biological Sciences*, *265*(1406), 1665–1673. https://doi.org/10.1098/rspb.1998.0486

Aubin, T., & Mathevon, N. (Eds.). (2020). *Coding Strategies in Vertebrate Acoustic Communication* (Vol. 7). Springer International Publishing. https://doi.org/10.1007/978-3-030-39200-0

Aylor, D. (1972). Noise Reduction by Vegetation and Ground. *The Journal of the Acoustical Society of America*, *51*(1B), 197–205. https://doi.org/10.1121/1.1912830

Brackenbury. (1979). Power Capabilities of the Avian Sound-Producing System. *Journal of Experimental Biology*, *78*(1), 163–166. https://doi.org/10.1242/jeb.78.1.163

Buxton, R. T., Lendrum, P. E., Crooks, K. R., & Wittemyer, G. (2018). Pairing camera traps and acoustic recorders to monitor the ecological impact of human disturbance. *Global Ecology and Conservation*, *16*, e00493. https://doi.org/10.1016/j.gecco.2018.e00493

Castro, I., Rosa, A. D., Priyadarshani, N., Bradbury, L., & Marsland, S. (2019). Experimental test of birdcall detection by autonomous recorder units and by human observers using broadcast. *Ecology and Evolution*, *9*(5), 2376–2397. https://doi.org/10.1002/ece3.4775

Dabelsteen, T., Larsen, O. N., & Pedersen, S. B. (1993). Habitat-induced degradation of sound signals: Quantifying the effects of communication sounds and bird location on blur ratio, excess attenuation, and signal-to-noise ratio in blackbird song. *The Journal of the Acoustical Society of America*, *93*(4), 2206–2220. https://doi.org/10.1121/1.406682

Darras, K., Batáry, P., Furnas, B. J., Grass, I., Mulyani, Y. A., & Tscharntke, T. (2019). Autonomous sound recording outperforms human observation for sampling birds: A systematic map and user guide. *Ecological Applications*, *29*(6). https://doi.org/10.1002/eap.1954

Darras, K., Furnas, B., Fitriawan, I., Mulyani, Y., & Tscharntke, T. (2018). Estimating bird detection distances in sound recordings for standardizing detection ranges and distance sampling. *Methods in Ecology and Evolution*, *9*(9), 1928–1938. https://doi.org/10.1111/2041-210X.13031

Darras, K., Pütz, P., Fahrurrozi, Rembold, K., & Tscharntke, T. (2016). Measuring sound detection spaces for acoustic animal sampling and monitoring. *Biological Conservation*, *201*, 29–37. https://doi.org/10.1016/j.biocon.2016.06.021

Dawson, D. K., & Efford, M. G. (2009). Bird population density estimated from acoustic signals. *Journal of Applied Ecology*, *46*(6), 1201–1209. https://doi.org/10.1111/j.1365-2664.2009.01731.x

Desjonquères, C., Gifford, T., & Linke, S. (2020). Passive acoustic monitoring as a potential tool to survey animal and ecosystem processes in freshwater environments. *Freshwater Biology*, *65*(1), 7–19. https://doi.org/10.1111/fwb.13356

Doser, J. W., Finley, A. O., Weed, A. S., & Zipkin, E. F. (2021). Integrating automated acoustic vocalization data and point count surveys for estimation of bird abundance. *Methods in Ecology and Evolution*, *n/a*(n/a). https://doi.org/10.1111/2041-210X.13578

Ellinger, N., & Hödl, W. (2003). Habitat acoustics of a neotropical lowland rainforest. *Bioacoustics*, *13*(3), 297–321. https://doi.org/10.1080/09524622.2003.9753503

Embleton, T. F. W. (1996). Tutorial on sound propagation outdoors. *The Journal of the Acoustical Society of America*, *100*(1), 31–48. https://doi.org/10.1121/1.415879

Fletcher, H., & Munson, W. A. (1933). Loudness, Its Definition, Measurement and Calculation*. *Bell System Technical Journal*, *12*(4), 377–430. https://doi.org/10.1002/j.1538-7305.1933.tb00403.x

Folliot, A., Haupert, S., Ducrettet, M., Sèbe, F., & Sueur, J. (2022). Using acoustics and artificial intelligence to monitor pollination by insects and tree use by woodpeckers. *Science of The Total Environment*, 155883. https://doi.org/10.1016/j.scitotenv.2022.155883

Fregosi, S., Harris, D. V., Matsumoto, H., Mellinger, D. K., Negretti, C., Moretti, D. J., Martin, S. W., Matsuyama, B., Dugan, P. J., & Klinck, H. (2020). Comparison of fin whale 20 Hz call detections by deep-water mobile autonomous and stationary recorders. *The Journal of the Acoustical Society of America*, *147*(2), 961–977. https://doi.org/10.1121/10.0000617

Gibb, R., Browning, E., Glover-Kapfer, P., & Jones, K. E. (2019). Emerging opportunities and challenges for passive acoustics in ecological assessment and monitoring. *Methods in Ecology and Evolution*, *10*(2), 169–185. https://doi.org/10.1111/2041-210X.13101

Grinfeder, E., Haupert, S., Ducrettet, M., Barlet, J., Reynet, M.-P., Sèbe, F., & Sueur, J. (2022). Soundscape dynamics of a cold protected forest: Dominance of aircraft noise. *Landscape Ecology*. https://doi.org/10.1007/s10980-021-01360-1

Haupert, S., Sèbe, F., & Sueur, J. (2022). Data and code from: Physics-based model to predict the acoustic detection distance of terrestrial autonomous recording units over the diel cycle and across seasons: insights from an Alpine and a Neotropical forest. *Zenodo,* https://doi.org/10.5281/zenodo.7229106

Hedley, R. W., Wilson, S. J., Yip, D. A., Li, K., & Bayne, E. M. (2021). Distance truncation via sound level for bioacoustic surveys in patchy habitat. *Bioacoustics*, *30*(3), 303–323. https://doi.org/10.1080/09524622.2020.1730240

Hill, A. P., Prince, P., Snaddon, J. L., Doncaster, C. P., & Rogers, A. (2019). AudioMoth: A low-cost acoustic device for monitoring biodiversity and the environment. *HardwareX*, *6*, e00073. https://doi.org/10.1016/j.ohx.2019.e00073

Kahl, S., Wood, C. M., Eibl, M., & Klinck, H. (2021). BirdNET: A deep learning solution for avian diversity monitoring. *Ecological Informatics*, *61*, 101236. https://doi.org/10.1016/j.ecoinf.2021.101236

Koper, N., Leston, L., Baker, T. M., Curry, C., & Rosa, P. (2016). Effects of ambient noise on detectability and localization of avian songs and tones by observers in grasslands. *Ecology and Evolution*, *6*(1), 245–255. https://doi.org/10.1002/ece3.1847

Ligges, U. (2016). *TuneR – Analysis of Music*.

MacLaren, A. R., Crump, P. S., Royle, J. A., & Forstner, M. R. J. (2018). *Observer-free experimental evaluation of habitat and distance effects on the detection of anuran and bird vocalizations*. 14.

Maechler, M., Rousseeuw, P., Croux, C., Todorov, V., Ruckstuhl, A., Salibian-Barrera, M., Verbeke, T., Koller, M., Conceicao, E. L., & Anna di Palma, M. (2022). *robustbase: Basic Robust Statistics. R package version 0.95-0.* http://robustbase.r-forge.r-projec t.org/

Marsh, A. H. (2001). IEC 61672, the new International Standard for sound level meters. *INTER-NOISE and NOISE-CON Congress and Conference Proceedings*, *2001*(3), 2240–2245.

Morton, E. S. (1975). Ecological Sources of Selection on Avian Sounds. *The American Naturalist*, *109*(965), 17–34.

Nemeth, E. (2004). Measuring the sound pressure level of the song of the screaming piha (Lipaugus vociferan): One of the loudest birds in the world? *Bioacoustics*, *14*(3), 225–228. https://doi.org/10.1080/09524622.2004.9753527

Newson, S. E., Bas, Y., Murray, A., & Gillings, S. (2017). Potential for coupling the monitoring of bush-crickets with established large-scale acoustic monitoring of bats. *Methods in Ecology and Evolution*, *8*(9), 1051–1062. https://doi.org/10.1111/2041-210X.12720

Pérez Granados, C., Bota, G., Albarracín, J., Giralt, D., & Traba, J. (2019). Cost-Effectiveness Assessment of Five Audio Recording Systems for Wildlife Monitoring: Differences between Recording Distances and Singing Direction. *Ardeola: Revista Ibérica de Ornitología*, *66*, 311–325. https://doi.org/10.13157/arla.66.2.2019.ra4

Pérez-Granados, C., Bota, G., Giralt, D., & Traba, J. (2018). A cost-effective protocol for monitoring birds using autonomous recording units: A case study with a night-time singing passerine. *Bird Study*, *65*(3), 338–345. https://doi.org/10.1080/00063657.2018.1511682

Pérez-Granados, C., & Traba, J. (2021). Estimating bird density using passive acoustic monitoring: A review of methods and suggestions for further research. *Ibis*, *163*(3), 765–783. https://doi.org/10.1111/ibi.12944

Price, M. A., Attenborough, K., & Heap, N. W. (1988). Sound attenuation through trees: Measurements and models. *The Journal of the Acoustical Society of America*, *84*(5), 1836–1844. https://doi.org/10.1121/1.397150

Priyadarshani, N., Marsland, S., & Castro, I. (2018). Automated birdsong recognition in complex acoustic environments: A review. *Journal of Avian Biology*, *49*(5), jav-01447. https://doi.org/10.1111/jav.01447

Rappaport, D. I., Royle, J. A., & Morton, D. C. (2020). Acoustic space occupancy: Combining ecoacoustics and lidar to model biodiversity variation and detection bias across heterogeneous landscapes. *Ecological Indicators*, *113*, 106172. https://doi.org/10.1016/j.ecolind.2020.106172

Rempel, R. S., Francis, C. M., Robinson, J. N., & Campbell, M. (2013). Comparison of audio recording system performance for detecting and monitoring songbirds. *Journal of Field Ornithology*, *84*(1), 86–97. https://doi.org/10.1111/jofo.12008

Römer, H. (2001). Ecological Constraints for Sound Communication: From Grasshoppers to Elephants. In F. G. Barth & A. Schmid (Eds.), *Ecology of Sensing* (pp. 59–77). Springer Berlin Heidelberg.

https://doi.org/10.1007/978-3-662-22644-5_4

Rossing, T. D. (Ed.). (2007). *Springer handbook of acoustics*. Springer.

Shaw, T., Müller, S., & Scherer-Lorenzen, M. (2021). Slope does not affect autonomous recorder detection shape: Considerations for acoustic monitoring in forested landscapes. *Bioacoustics*, 1–22. https://doi.org/10.1080/09524622.2021.1925590

Shonfield, J., & Bayne, E. M. (2017). Autonomous recording units in avian ecological research: Current use and future applications. *Avian Conservation and Ecology*, 13.

Simons, T. R., Alldredge, M. W., Pollock, K. H., & Wettroth, J. M. (2007). Experimental Analysis of the Auditory Detection Process on Avian Point Counts. *Análisis Experimentales Del Proceso de Detección Auditiva En Puntos de Conteo de Aves.*, *124*(3), 986–1000. https://doi.org/10.1642/0004-8038(2007)124[986:EAOTAD]2.0.CO;2

Sólymos, P., Matsuoka, S. M., Bayne, E. M., Lele, S. R., Fontaine, P., Cumming, S. G., Stralberg, D., Schmiegelow, F. K. A., & Song, S. J. (2013). Calibrating indices of avian density from non-standardized survey data: Making the most of a messy situation. *Methods in Ecology and Evolution*, *4*(11), 1047–1058. https://doi.org/10.1111/2041-210X.12106

Sousa-Lima, R. S., Norris, T. F., Oswald, J. N., & Fernandes, D. P. (2013). A review and inventory of fixed autonomous recorders for passive acoustic monitoring of marine mammals. *Aquatic Mammals*, *39*(1), 23–54.

Sueur, J., Aubin, T., & Simonis, C. (2008). Seewave, a Free Modular Tool for Sound Analysis and Synthesis. *Bioacoustics*, *18*(2), 213–226. https://doi.org/10.1080/09524622.2008.9753600

Sugai, L. S. M., Desjonquères, C., Silva, T. S. F., & Llusia, D. (2020). A roadmap for survey designs in terrestrial acoustic monitoring. *Remote Sensing in Ecology and Conservation*, *6*(3), 220–235. https://doi.org/10.1002/rse2.131

Sugai, L. S. M., Silva, T. S. F., Ribeiro, J. W., & Llusia, D. (2019). Terrestrial Passive Acoustic Monitoring: Review and Perspectives. *BioScience*, *69*(1), 15–25. https://doi.org/10.1093/biosci/biy147

Szabo, T. L. (1994). Time domain wave equations for lossy media obeying a frequency power law. *The Journal of the Acoustical Society of America*, *96*(1), 491–500. https://doi.org/10.1121/1.410434

Tarrero, A. I., Martín, M. A., González, J., Machimbarrena, M., & Jacobsen, F. (2008). Sound propagation in forests: A comparison of experimental results and values predicted by the Nord 2000 model. *Applied Acoustics*, *69*(7), 662–671. https://doi.org/10.1016/j.apacoust.2007.01.007

TEAM, R.Core. (2020). *R: A language and environment for statistical computing.*

Thomas, A., Speldewinde, P., Roberts, J. D., Burbidge, A. H., & Comer, S. (2020). If a bird calls, will we detect it? Factors that can influence the detectability of calls on automated recording units in field conditions. *Emu - Austral Ornithology*, *120*(3), 239–248. https://doi.org/10.1080/01584197.2020.1787848

Turgeon, P., Van Wilgenburg, S., & Drake, K. (2017). Microphone variability and degradation: Implications for monitoring programs employing autonomous recording units. *Avian Conservation and Ecology*, *12*, 9. https://doi.org/10.5751/ACE-00958-120109

Ulloa, J. S., Aubin, T., Llusia, D., Bouveyron, C., & Sueur, J. (2018). Estimating animal acoustic diversity in tropical environments using unsupervised multiresolution analysis. *Ecological Indicators*, *90*, 346–355. https://doi.org/10.1016/j.ecolind.2018.03.026

Ulloa, J. S., Haupert, S., Latorre, J. F., Aubin, T., & Sueur, J. (2021). scikit-maad: An open-source and modular toolbox for quantitative soundscape analysis in Python. *Methods in Ecology and Evolution*, 2041-210X.13711. https://doi.org/10.1111/2041-210X.13711

Van Wilgenburg, S. L., Sólymos, P., Kardynal, K. J., & Frey, M. D. (2017). Paired sampling standardizes point count data from humans and acoustic recorders. *Avian Conservation and Ecology*, *12*(1), art13. https://doi.org/10.5751/ACE-00975-120113

Wickham, H., Averick, M., Bryan, J., Chang, W., McGowan, L., François, R., Grolemund, G., Hayes, A., Henry, L., Hester, J., Kuhn, M., Pedersen, T., Miller, E., Bache, S., Müller, K., Ooms, J., Robinson, D., Seidel, D., Spinu, V., … Yutani, H. (2019). Welcome to the Tidyverse. *The Journal of Open Source Software*, *4*, 1686. https://doi.org/10.21105/joss.01686

Wiener, F. M., & Keast, D. N. (1959). Experimental Study of the Propagation of Sound over Ground. *The Journal of the Acoustical Society of America*, *31*(6), 724–733. https://doi.org/10.1121/1.1907778

Yip, D. A., & Bayne, E. M. (2017). Sound attenuation in forest and roadside environments: Implications for avian point-count surveys. *The Condor*, 12.

Yip, D. A., Knight, E. C., Haave-Audet, E., Wilson, S. J., Charchuk, C., Scott, C. D., Sólymos, P., & Bayne, E. M. (2020). Sound level measurements from audio recordings provide objective distance estimates for distance sampling wildlife populations. *Remote Sensing in Ecology and Conservation*, *6*(3), 301–315. https://doi.org/10.1002/rse2.118

Yip, D., Leston, L., Bayne, E., Sólymos, P., & Grover, A. (2017). Experimentally derived detection distances from audio recordings and human observers enable integrated analysis of point count data. *Avian Conservation and Ecology*, *12*(1). https://doi.org/10.5751/ACE-00997-120111

## Authors' Contributions

Tabular author's contributions modified after the CRediT system (https://credit.niso.org/). All authors contributed critically to the drafts and gave final approval for publication.

| Contribution | SH | FS | JS |
|---|---|---|---|
| Conceptualization | ■ | ■ | ■ |
| Investigation | ■ | ■ | |
| Data curation | ■ | | |
| Formal analysis | ■ | | |
| Visualization | ■ | | |
| Software | ■ | | |
| Writing – original draft | ■ | ■ | ■ |
| Writing – review and editing | ■ | ■ | ■ |
| Supervision | | | ■ |

## Acknowledgements

We would like to warmly thank Pierre-Michel Forget from MECADEV (MNHN) for his expertise in tropical ecology, and Nina Marchand, Philippe Gaucher, Elodie Courtois and Patrick Châtelet from LEESIA (CNRS) for their help during our stay at the Nouragues research field station. We would like to thank Marie-Pierre Reynet and Julien Barlet from the Parc Naturel Regional du Haut-Jura for their help during our experiment in the Risoux forest in Jura. Finally, we would like to thank Patrick Quartier from the LAM (CNRS) for characterizing the frequency response of the Song Meter SM4 (Wildlife Acoustics, Maynard, MA, USA) in the LAM's anechoic chamber.

## Conflict of interest

The authors have no conflict of interest to declare.

## Funding

This work was funded by the Nouragues research field station (managed by CNRS), which benefits from Investissement d'Avenir grants managed by the Agence Nationale de la Recherche (AnaEE France ANR-11-INBS-0001; Labex CEBA ANR-10-LABX-25-01). This study was also funded by a grant from the Parc Naturel Régional du Haut-Jura (Région Bourgogne-Franche-Comté, Région Auvergne-Rhône-Alpes, Direction Régionale de l'Environnement, de l'Aménagement et du Logement de Bourgogne-Franche-Comté).

## Data Availability

The data and codes supporting the results are available at https://github.com/shaupert/HAUPERT_MEE_2022 and are referenced on Zenodo (Haupert et al/, 2022; DOI: https://doi.org/10.5281/zenodo.7229106)

**Table legends**

Table 1. Main experimental parameters. Equipment settings (ARU, sound level meter, loudspeaker), environmental variables and spatial sampling used in the two forest sites. The parameters slightly differ between the sites due different sampling protocols used for long-term monitoring programs.

Table 2: Habitat attenuation coefficients $a_0$ depending on the habitat in the neotropical rainforest (French Guiana) and in the Alpine coniferous forest (Jura), the recorder (reference sound level meter and ARU), and the reference distance $r_0$ (m). The unit of the coefficient $a_0$ is dB/kHz/m.

Table 1

|  | **Neotropical rainforest (French Guiana)** | **Alpine coniferous forest (Jura)** |
|---|---|---|
| **SM4 audio recording settings** |  |  |
| Sampling frequency | 48,000 Hz | 44,100 Hz |
| Pre-amplifier gain | 26 dB | 26 dB |
| Gain | 4 dB | 16 dB |
| Microphone sensitivity | –35 dBV | –35 dBV |
| Filter | No | No |
| Height | 1.5 m | 2.5 m |
| **Sound level meter settings** |  |  |
| Sampling frequency | 48,000 Hz | 48,000 Hz |
| Pre-amplifier gain | 12 dB | 12 dB |
| Gain | None | None |
| Microphone sensitivity | –29.12 dBV | –29.12 dBV |
| Filter | No | No |
| Height | 1.5 m | 2.5 m |
| **Loudspeaker settings** |  |  |
| Output sound level (SPL) | 83 dB | 78 dB |
| Signal type | White noise | White noise |
| Bandwidth | 70 Hz – 20 kHz | 70 Hz – 20 kHz |
| **Weather** |  |  |
| Temperature | 24 °C | 17 °C |
| Relative humidity | 87 % | 67 % |
| Atmospheric pressure | 101,340 Pa | 87,999 Pa |
| **Transect** |  |  |
| Positions [m] | 1; 10; 20; 30; 40; 50; 60; 70; 80; 90; 100 | 1; 2; 5; 10; 20; 30; 40; 50; 60; 70; 80; 90; 100 |
| Altitude | 69 m +/– 5m | 1,210 m +/– 4m |

Table 2

| | Neotropical rainforest (French Guiana) | | Alpine coniferous forest (Jura) | |
|---|---|---|---|---|
| $r_0$ | Reference | ARU | Reference | ARU |
| 10 m | 0.018 ($R^2 = 0.7$) | 0.012 ($R^2 = 0.6$) | 0.014 ($R^2 = 0.6$) | 0.028 ($R^2 = 0.9$) |
| 20 m | 0.019 ($R^2 = 0.7$) | 0.010 ($R^2 = 0.3$) | 0.033 ($R^2 = 0.8$) | 0.033 ($R^2 = 0.8$) |
| 30 m | 0.018 ($R^2 = 0.8$) | 0.010 ($R^2 = 0.5$) | 0.024 ($R^2 = 0.7$) | 0.021 ($R^2 = 0.7$) |
| 40 m | 0.019 ($R^2 = 0.6$) | *p > 0.001* | 0.017 ($R^2 = 0.5$) | 0.014 ($R^2 = 0.7$) |
| 50 m | *p > 0.001* | *p > 0.001* | 0.013 ($R^2 = 0.5$) | *p > 0.001* |
| $a_0$ mean | 0.019 | 0.011 | 0.020 | 0.024 |
| $a_0$ std | 0.001 | 0.001 | 0.008 | 0.008 |